\documentclass[%
reprint,
showpacs,preprintnumbers,
nofootinbib,
amsmath,amssymb,
aps,
prd,
floatfix,
]{revtex4-1}
\usepackage{graphicx}
\usepackage{dcolumn}
\usepackage{bm}
\usepackage{url}
\usepackage[ddmmyy,24hr]{datetime}
\usepackage{color}

\newcommand{\nua}[1]{\ensuremath{\rlap
           {\kern-2.5pt\ensuremath
           {\overset{\scriptscriptstyle(-)}{\phantom{\nu}}}}
           {\ensuremath{{\nu}_{#1}}}}}
\begin{document}

\preprint{\begin{tabular}{l}
\texttt{arXiv:1308.5288 [hep-ph]}
\\
\textbf{Phys. Rev. D 88, 073008 (2013)}
\end{tabular}}

\title{Pragmatic View of Short-Baseline Neutrino Oscillations}

\author{C. Giunti}
\affiliation{INFN, Sezione di Torino, Via P. Giuria 1, I--10125 Torino, Italy}

\author{M. Laveder}
\affiliation{Dipartimento di Fisica e Astronomia ``G. Galilei'', Universit\`a di Padova,
and
INFN, Sezione di Padova,
Via F. Marzolo 8, I--35131 Padova, Italy}

\author{Y.F. Li}
\affiliation{Institute of High Energy Physics,
Chinese Academy of Sciences, Beijing 100049, China}

\author{H.W. Long}
\affiliation{Department of Modern Physics, University of Science and
Technology of China, Hefei, Anhui 230026, China}

\date{\dayofweekname{\day}{\month}{\year} \ddmmyydate\today, \currenttime}

\begin{abstract}
We present the results of global analyses of short-baseline
neutrino oscillation data in 3+1, 3+2 and 3+1+1 neutrino mixing schemes.
We show that the data do not allow us to abandon the simplest 3+1 scheme
in favor of the more complex 3+2 and 3+1+1 schemes.
We present the allowed region in the 3+1 parameter space,
which is located at $\Delta{m}^2_{41}$
between
$0.82$
and
$2.19$
$\text{eV}^2$
at $3\sigma$.
The case of no oscillations is disfavored by about $6\sigma$,
which decreases dramatically to about $2\sigma$
if the LSND data are not considered.
Hence, new high-precision experiments are needed to check the LSND signal.
\end{abstract}


\maketitle

The possibility of short-baseline neutrino oscillations
due to the existence of one or more sterile neutrinos at the eV scale
is a hot topic in current neutrino physics
(see \cite{0704.1800,1204.5379,1302.1102,1303.6912}).
Besides the intrinsic interest in determining the existence of new phenomena and particles,
the existence and properties of sterile neutrinos and active-sterile mixing
could shed light on the physics beyond the Standard Model
(see \cite{hep-ph/0111326,hep-ph/0603118}).
The existence of light sterile neutrinos is also very important for astrophysics
(see \cite{1206.6231})
and cosmology
(see \cite{1301.7102,1307.0637}),
and the recent first Planck results \cite{1303.5076}
have generated interesting studies on the implications of
cosmological data for light sterile neutrinos
\cite{1303.5368,1303.6267,1304.5981,1304.6217,1307.2904,1307.7715,1308.3255}.

In this paper we extend the analysis
of short-baseline electron neutrino and antineutrino disappearance data
presented in Ref.~\cite{1210.5715}
by taking into account also the more controversial
indication of the
LSND \cite{Aguilar:2001ty}
experiment
in favor of short-baseline
$\bar\nu_{\mu}\to\bar\nu_{e}$
transitions
and the recent ambiguous results of the
MiniBooNE \cite{1303.2588}
experiment.
We consider
3+1 and 3+2 extensions of the Standard Model
in which there are, respectively, one or two sterile neutrinos at the eV scale\footnote{
Our approach does not exclude the possible existence of more sterile neutrinos
as long as the effects of their mixing with the active neutrino is negligible
for the analysis of current short-baseline neutrino oscillation experiments.
Heavier sterile neutrinos with masses at the keV scale have been considered as
candidates for warm dark matter
(see~\cite{0901.0011,0906.2968,1303.4758,1306.4954}).
Very light sterile neutrinos with masses smaller than about 0.1 eV
can induce observable effects
in solar neutrino experiments
\cite{0902.1310,1012.5627}
and in long-baseline reactor experiments
\cite{0809.5076,1308.2823,1308.5880,1308.6218}.
}
which generate short-baseline oscillations
\cite{1103.4570,1107.1452,1109.4033,1111.1069,1207.4765,1207.6515,1210.5715,1212.3805,1302.6720,1303.3011}.
In the 3+1 scheme
electron and muon neutrino and antineutrino appearance and disappearance
in short-baseline experiments
depend on:
\begin{enumerate}

\renewcommand{\labelenumi}{\theenumi)}
\renewcommand{\theenumi}{\Alph{enumi}}

\item
One neutrino squared-mass difference,
$\Delta{m}^2_{41} = m_{4}^2 - m_{1}^2$,
where $m_{k}$ is the mass of the massive neutrino $\nu_{k}$
and
$\Delta{m}^2_{21} \ll \Delta{m}^2_{31} \ll \Delta{m}^2_{41} \sim 1 \, \text{eV}^2$
in order to accommodate the observed oscillations of
solar, reactor, atmospheric and accelerator neutrinos
in the standard framework of three-neutrino mixing
(see \cite{Giunti-Kim-2007,0704.1800}).
The probability of
$\nua{\alpha}\to\nua{\beta}$
transitions has the two-neutrino-like form
\begin{equation}
P_{\nua{\alpha}\to\nua{\beta}}
=
\delta_{\alpha\beta}
-
4 |U_{\alpha4}|^2 \left( \delta_{\alpha\beta} - |U_{\beta4}|^2 \right)
\sin^2\!\left( \dfrac{\Delta{m}^2_{41}L}{4E} \right)
\,,
\label{pab}
\end{equation}
where $U$ is the mixing matrix,
$L$ is the source-detector distance,
and $E$ is the neutrino energy.

\item
$|U_{e4}|^2$ and $|U_{\mu4}|^2$,
which
determine the amplitude
$\sin^22\vartheta_{e\mu} = 4 |U_{e4}|^2 |U_{\mu4}|^2$
of
$\nua{\mu}\to\nua{e}$
transitions,
the amplitude
$\sin^22\vartheta_{ee} = 4 |U_{e4}|^2 \left( 1 - |U_{e4}|^2 \right)$
of
$\nua{e}$
disappearance,
and
the amplitude
$\sin^22\vartheta_{\mu\mu} = 4 |U_{\mu4}|^2 \left( 1 - |U_{\mu4}|^2 \right)$
of
$\nua{\mu}$
disappearance.

\end{enumerate}

Since the oscillation probabilities of neutrinos and antineutrinos are related by
a complex conjugation of the elements of the mixing matrix
(see \cite{Giunti-Kim-2007}),
the probabilities of short-baseline
$\nu_{\mu}\to\nu_{e}$ and $\bar\nu_{\mu}\to\bar\nu_{e}$
transitions are equal.
Hence,
the 3+1 scheme cannot explain a possible CP-violating difference of
$\nu_{\mu}\to\nu_{e}$ and $\bar\nu_{\mu}\to\bar\nu_{e}$
transitions in short-baseline experiments.
In order to allow this possibility,
one must consider a 3+2 scheme,
in which, there are four additional effective mixing parameters in short-baseline experiments:

\begin{enumerate}

\setcounter{enumi}{2}

\renewcommand{\labelenumi}{\theenumi)}
\renewcommand{\theenumi}{\Alph{enumi}}

\item
$\Delta{m}^2_{51}$,
which is conventionally assumed $\geq\Delta{m}^2_{41}$.

\item
$|U_{e5}|^2$ and $|U_{\mu5}|^2$.

\item
$\eta = \text{arg}\left[U_{e4}^*U_{\mu4}U_{e5}U_{\mu5}^*\right]$.
Since this complex phase appears with different signs in
$\nu_{\mu}\to\nu_{e}$ and $\bar\nu_{\mu}\to\bar\nu_{e}$
transitions, it can generate measurable CP violations.

\end{enumerate}

\begin{figure*}[t]
\null
\hfill
\includegraphics*[width=0.49\linewidth]{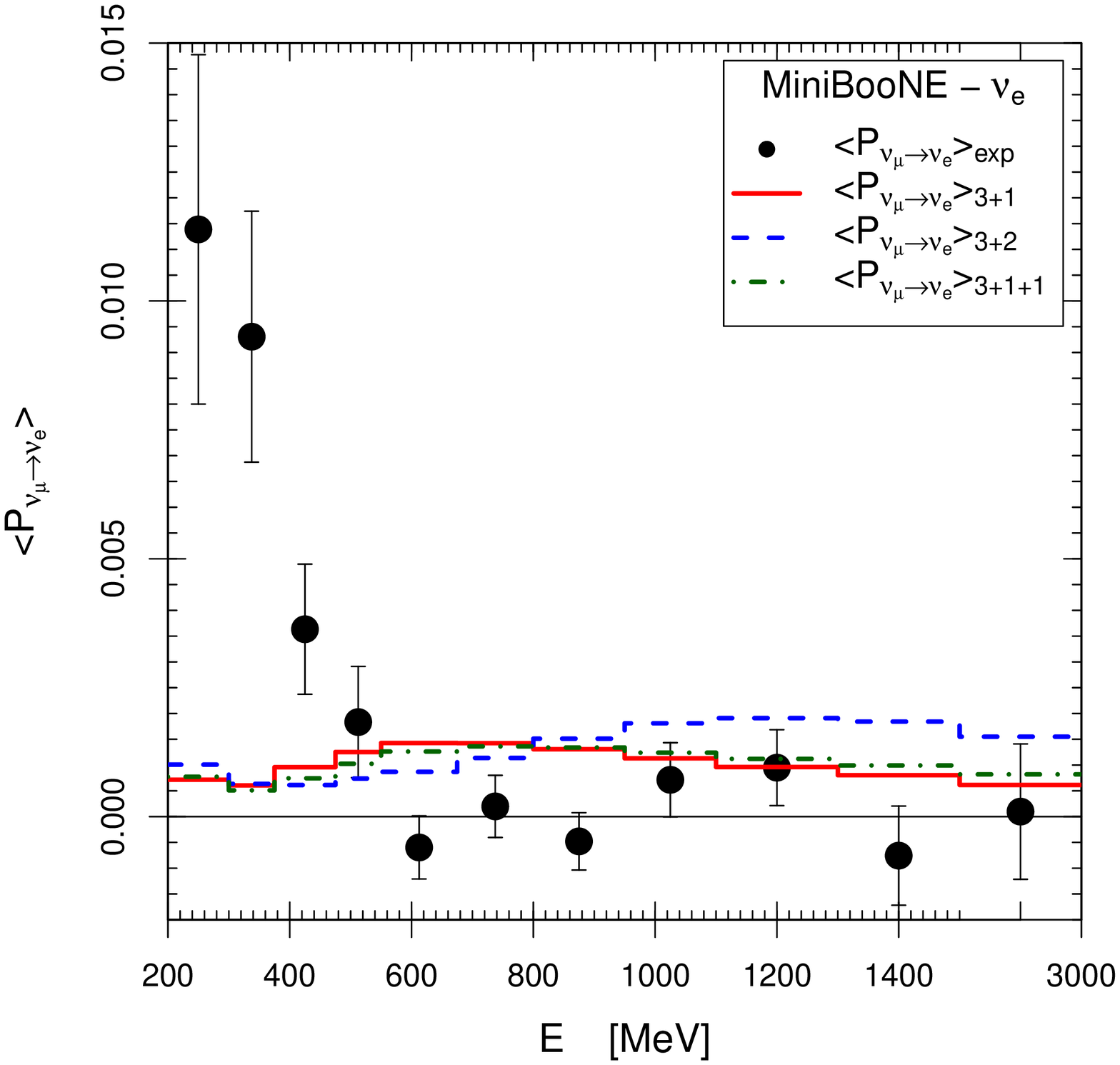}
\hfill
\includegraphics*[width=0.49\linewidth]{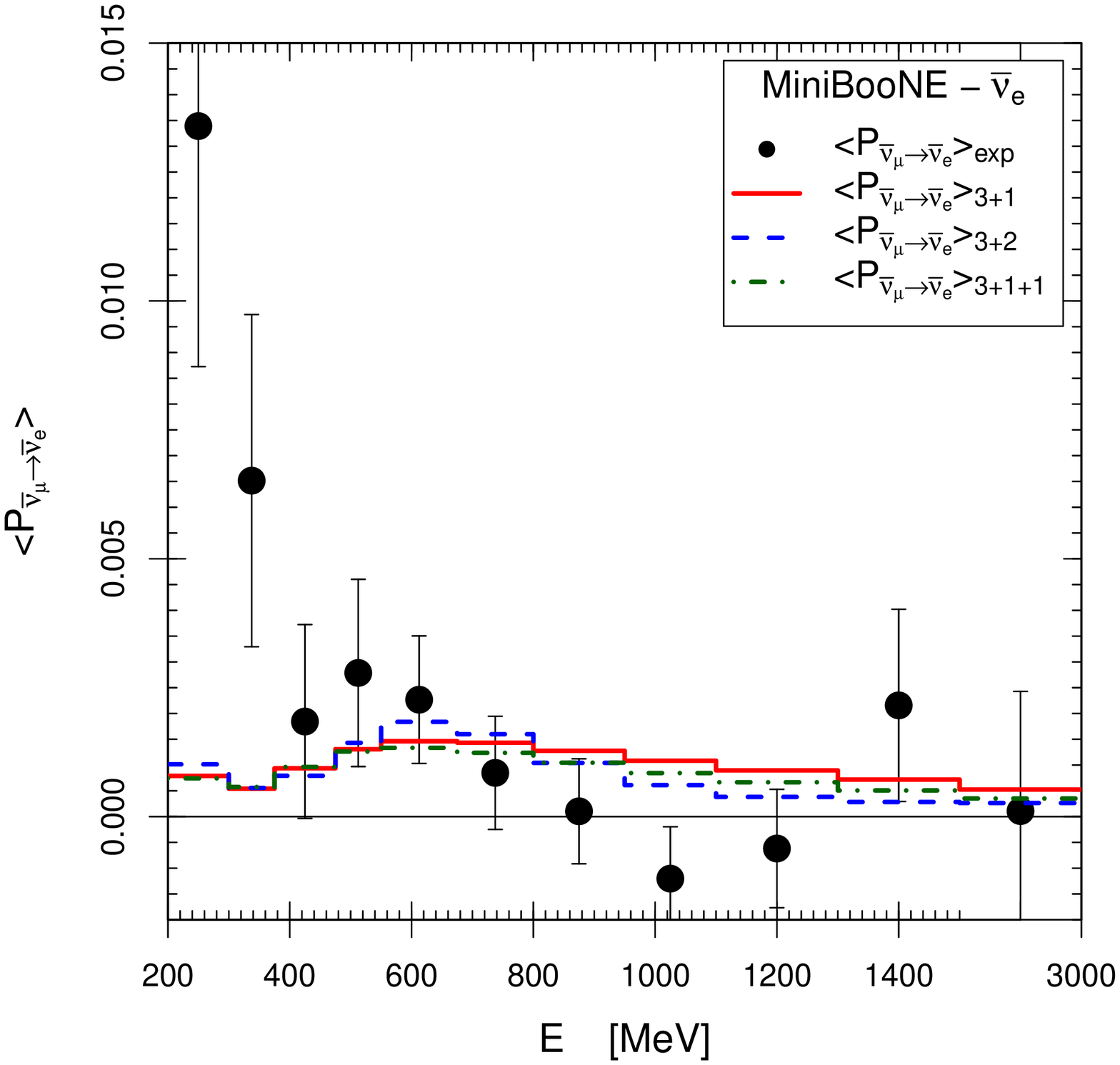}
\hfill
\null
\caption{ \label{fig:MB}
Averaged transition probability in MiniBooNE bins corresponding to the best-fit values of the
oscillation parameters
in the 3+1-LOW, 3+2-LOW and 3+1+1-LOW fits
(see Tabs.~\protect\ref{tab:3p1} and \protect\ref{tab:3p2})
compared with the experimental data.
}
\end{figure*}

In the analysis of short-baseline data,
we consider the following three groups of experiments:

\begin{enumerate}

\item
The
$\nua{\mu}\to\nua{e}$
appearance data of the
LSND \cite{Aguilar:2001ty},
MiniBooNE \cite{1303.2588},
BNL-E776 \cite{Borodovsky:1992pn},
KARMEN \cite{Armbruster:2002mp},
NOMAD \cite{Astier:2003gs},
ICARUS \cite{1307.4699}
and
OPERA \cite{1303.3953}
experiments.

\item
The
$\nua{e}$
disappearance data described in Ref.~\cite{1210.5715},
which take into account the
reactor
\cite{1101.2663,1101.2755,1106.0687}
and
Gallium
\cite{nucl-ex/0512041,Laveder:2007zz,hep-ph/0610352,0711.4222,1006.3244}
anomalies.

\item
The constraints on
$\nua{\mu}$
disappearance obtained from
the data of the
CDHSW experiment \cite{Dydak:1983zq},
from the analysis \cite{Maltoni:2007zf} of
the data of
atmospheric neutrino oscillation experiments\footnote{
We do not consider the IceCube data which could give a marginal contribution
\cite{1206.6903,1307.6824},
because the analysis is too complicated and subject to large uncertainties.
},
from the analysis \cite{1109.4033} of the
MINOS neutral-current data \cite{Adamson:2011ku}
and from the analysis of the
SciBooNE-MiniBooNE
neutrino \cite{Mahn:2011ea} and antineutrino \cite{Cheng:2012yy} data.

\end{enumerate}

With respect to the analysis presented in Ref.~\cite{1302.6720},
we have added the new constraints on
$\nu_{\mu}\to\nu_{e}$
transitions obtained in the
ICARUS \cite{1307.4699}
and
OPERA \cite{1303.3953}
experiments.
Following Ref.~\cite{1303.3011},
we
also added the constraints on
$\nu_{\mu}\to\nu_{e}$ and $\bar\nu_{\mu}\to\bar\nu_{e}$
appearance
obtained in the old
BNL-E776 \cite{Borodovsky:1992pn}
experiment,
which give a small contribution at $\Delta{m}^2$'s larger than about $2 \, \text{eV}^2$,
and
the subleading effect of background disappearance in the analysis of
MiniBooNE \cite{1303.2588}
data.

\begin{table}[t]
\begin{ruledtabular}
\begin{tabular}{cccccc}
		&				&LOW							&HIG							&noMB							&noLSND							\\
\hline
No		&$\chi^{2}$			&339.2		&308.0		&283.2		&286.7		\\
Osc.		&NDF				&259		&253		&221		&255		\\
		&GoF				&0.06\%	& 1\%	&0.3\%	& 8\%	\\
\hline
3+1		&$\chi^{2}_{\text{min}}$	&291.7		&261.8		&236.1		&278.4		\\
Osc.		&NDF				&256		&250		&218		&252		\\
		&GoF				& 6\%		&29\%		&19\%		&12\%		\\
		&$\Delta{m}^2_{41}[\text{eV}^2]$&1.6		&1.6		&1.6		&1.7		\\
		&$|U_{e4}|^2$			&0.033		&0.03		&0.03		&0.024		\\
		&$|U_{\mu4}|^2$			&0.012		&0.013		&0.014		&0.0073		\\
		&$\sin^22\vartheta_{e\mu}$	&0.0016		&0.0015		&0.0017		&0.0007		\\
		&$\sin^22\vartheta_{ee}$	&0.13		&0.11		&0.12		&0.093		\\
		&$\sin^22\vartheta_{\mu\mu}$	&0.048		&0.049		&0.054		&0.03		\\
\hline
	&$(\chi^{2}_{\text{min}})_{\text{APP}}$	&99.3		&77.0		&50.9		&91.8		\\
	&$(\chi^{2}_{\text{min}})_{\text{DIS}}$	&180.1		&180.1		&180.1		&180.1		\\
		&$\Delta\chi^{2}_{\text{PG}}$	&12.7		&4.8		&5.1		&6.4		\\
		&$\text{NDF}_{\text{PG}}$	&2		&2		&2		&2		\\
		&$\text{GoF}_{\text{PG}}$	&0.2\%	& 9\%	& 8\%	& 4\%	\\
\hline
		&$\text{p-val}_{\text{No Osc.}}$&$3
					\times10^{-10}$	&$5
											\times10^{-10}$			&$3
																			\times10^{-10}$		&$4
																										\times10^{-2}$		\\
		&$n\sigma_{\text{No Osc.}}$	&$6.3\sigma$	&$6.2\sigma$	&$6.3\sigma$	&$2.1\sigma$	\\
\end{tabular}
\end{ruledtabular}
\caption{ \label{tab:3p1}
Results of the fit of short-baseline data
taking into account all MiniBooNE data (LOW),
only the MiniBooNE data above 475 MeV (HIG),
without MiniBooNE data (noMB)
and without LSND data (noLSND).
The results of the fit without neutrino oscillations are given in the first three lines,
whereas the other lines refer to the 3+1 fit.
We list the $\chi^{2}$,
the number of degrees of freedom (NDF),
the goodness-of-fit (GoF),
the best-fit values of the 3+1 oscillation parameters
and
the quantities relevant for the appearance-disappearance (APP-DIS) parameter goodness-of-fit (PG)
\protect\cite{hep-ph/0304176}.
In the last three lines we give
the p-value ($\text{p-val}_{\text{No Osc.}}$)
and
the corresponding number of excluding $\sigma$'s ($n\sigma_{\text{No Osc.}}$) of the no-oscillation case.
}
\end{table}

\begin{figure}[t]
\begin{center}
\includegraphics*[width=\linewidth]{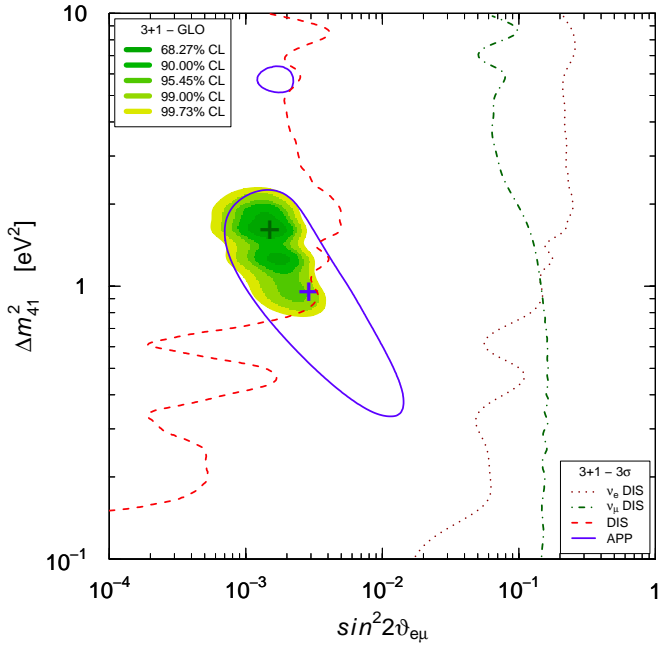}
\end{center}
\caption{ \label{fig:app}
Allowed region in the
$\sin^{2}2\vartheta_{e\mu}$--$\Delta{m}^{2}_{41}$ plane
in the global (GLO) 3+1-HIG fit
of short-baseline neutrino oscillation data
compared with the $3\sigma$ allowed regions
obtained from
$\protect\nua{\mu}\to\protect\nua{e}$
short-baseline appearance data (APP; inside the solid blue curves)
and the $3\sigma$ constraints obtained from
$\protect\nua{e}$
short-baseline disappearance data ($\nu_{e}$ DIS; left of the dotted dark-red curve),
$\protect\nua{\mu}$
short-baseline disappearance data ($\nu_{\mu}$ DIS; left of the dash-dotted dark-green curve)
and the
combined short-baseline disappearance data (DIS; left of the dashed red curve).
The best-fit points of the GLO and APP fits are indicated by crosses.
}
\end{figure}

\begin{figure}[t]
\begin{center}
\includegraphics*[width=\linewidth]{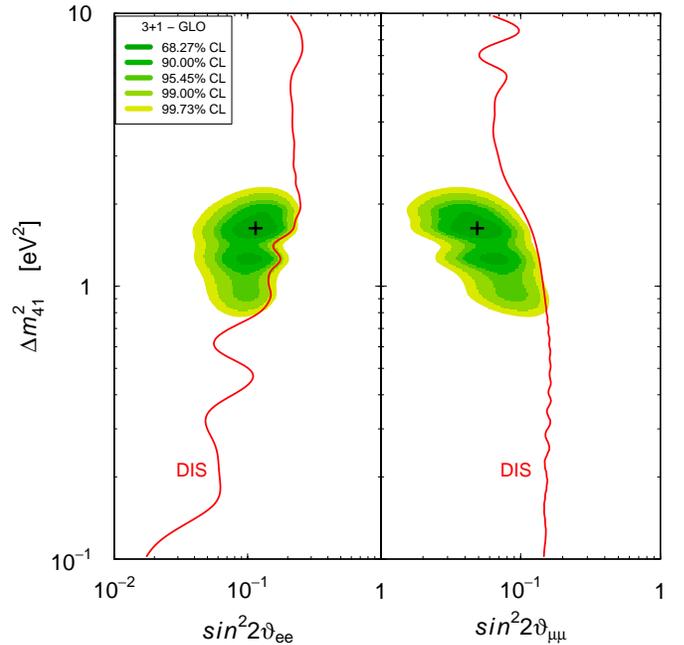}
\end{center}
\caption{ \label{fig:dis}
Allowed regions in the
$\sin^{2}2\vartheta_{ee}$--$\Delta{m}^{2}_{41}$
and
$\sin^{2}2\vartheta_{\mu\mu}$--$\Delta{m}^{2}_{41}$
planes
in the global (GLO) 3+1-HIG fit
of short-baseline neutrino oscillation data
compared with the $3\sigma$ constraints obtained from
$\protect\nua{e}$
short-baseline disappearance data (solid red DIS curve in the left panel),
$\protect\nua{\mu}$
short-baseline disappearance data (solid red DIS curve in the right panel).
The best-fit point of the GLO fit is indicated by crosses.
}
\end{figure}

Table~\ref{tab:3p1}
shows the results of the 3+1-LOW fit
of all the data above,
including the three low-energy bins of the MiniBooNE experiment
whose excess with respect to the background
is widely considered to be anomalous
because it is at odds with neutrino oscillations
\cite{1109.4033,1111.1069}.
We have considered also a 3+1-HIG fit
of MiniBooNE data
without the three anomalous low-energy bins.
From Tab.~\ref{tab:3p1},
one can see that in both cases the oscillation fit of the data
is much better than the no-oscillation fit,
which has a disastrous p-value
and is excluded in both cases at about $6\sigma$.
Although the best-fit values of the oscillation parameters are similar
in the
3+1-LOW and 3+1-HIG fits,
the goodness-of-fit of the 3+1-LOW case is significantly lower
and the appearance-disappearance parameter goodness-of-fit is much lower.
This result confirms the fact that the MiniBooNE low-energy anomaly
is incompatible with neutrino oscillations,
because it would require a small value of $\Delta{m}^2_{41}$
and a large value of $\sin^22\vartheta_{e\mu}$
\cite{1109.4033,1111.1069},
which are excluded by the data of other experiments.
Indeed, one can see from Fig.~\ref{fig:MB}
that the best-fit 3+1-LOW averaged transition probability is far from fitting
the three anomalous low-energy bins of MiniBooNE neutrino and antineutrino data.
Therefore,
we think that it is very likely that the MiniBooNE low-energy anomaly
has an explanation which is different from neutrino oscillations\footnote{
The interesting possibility of reconciling the low--energy anomalous data with neutrino oscillations
through energy reconstruction effects proposed in \cite{Martini:2012fa,Martini:2012uc}
still needs a detailed study.
}
and the 3+1-HIG fit is more reliable than the 3+1-LOW fit.
Moreover,
the fact that both the
global goodness-of-fit
and
the appearance-disappearance parameter goodness-of-fit
are acceptable
in the 3+1-HIG fit
tells us that the fit is reliable.
Hence, in Figs.~\ref{fig:app} and \ref{fig:dis}
we present the allowed regions in the
$\sin^{2}2\vartheta_{e\mu}$--$\Delta{m}^{2}_{41}$,
$\sin^{2}2\vartheta_{ee}$--$\Delta{m}^{2}_{41}$ and
$\sin^{2}2\vartheta_{\mu\mu}$--$\Delta{m}^{2}_{41}$
planes,
which are relevant, respectively, for
$\nua{\mu}\to\nua{e}$ appearance,
$\nua{e}$ disappearance and
$\nua{\mu}$ disappearance
searches.
One can see that the allowed region is well defined,
with
\begin{equation}
0.82
<
\Delta{m}^2_{41}
<
2.19
\,
\text{eV}^2
\quad
(3\sigma)
\,.
\label{dm2}
\end{equation}
Figure~\ref{fig:app}
shows also the region allowed by $\nua{\mu}\to\nua{e}$ appearance data
and
the constraints on $\sin^{2}2\vartheta_{e\mu}$ from
$\nua{e}$ disappearance and
$\nua{\mu}$ disappearance data.
One can see that the combined disappearance constraint
excludes a large part of the region allowed by $\nua{\mu}\to\nua{e}$ appearance data,
leading to the well-known
appearance-disappearance tension
\cite{1103.4570,1107.1452,1109.4033,1111.1069,1207.4765,1207.6515,1302.6720,1303.3011}
quantified by the parameter goodness-of-fit in Tab.~\ref{tab:3p1}.
With respect to the results presented in Refs.~\cite{1107.1452,1109.4033,1111.1069,1207.6515,1302.6720},
the region at $\Delta{m}^2_{41} \simeq 6 \, \text{eV}^2$
which is allowed by
$\nua{\mu}\to\nua{e}$ appearance data
is not allowed any more (at $3\sigma$) by the global fit,
mainly because of the old BNL-E776 data,
which we included following the wise approach of Ref.~\cite{1303.3011}.
This is consistent with the cosmological exclusion of this region
\cite{1207.6515,1302.6720}.

It is interesting to investigate what is the
impact of the MiniBooNE experiment
towards the test of the LSND signal.
With this aim we performed two additional 3+1 fits:
a 3+1-noMB fit without MiniBooNE data
and
a 3+1-noLSND fit without LSND data.
From Tab.~\ref{tab:3p1}
one can see that the results of the
3+1-noMB fit are similar to those of the
3+1-HIG fit
and the exclusion of the case of no-oscillations remains at the level of $6\sigma$.
On the other hand,
in the 3+1-noLSND fit
the exclusion of the case of no-oscillations drops dramatically to\footnote{
This is due to the fact that without LSND the main indication in favor of short-baseline oscillations
is given by the reactor
\cite{1101.2663,1101.2755,1106.0687}
and
Gallium
\cite{1006.3244}
anomalies.
In fact,
the
$2.1\sigma$
exclusion
($
\Delta\chi^2 / \text{NDF}
=
8.3
/
3
$)
of the case of no-oscillations in the global fit
of short-baseline neutrino oscillation data
without LSND
is comparable with the
$2.7\sigma$
exclusion
($
\Delta\chi^2 / \text{NDF}
=
10.1
/
2
$)
that we obtain from the analysis of $\nua{e}$
short-baseline disappearance data alone \cite{1210.5715}.
}
$2.1\sigma$.
Therefore,
it is evident that the LSND experiment is still crucial for the indication in favor of short-baseline
$\bar\nu_{\mu}\to\bar\nu_{e}$
transitions
and the MiniBooNE experiment has been rather inconclusive.

\begin{table}[t]
\begin{ruledtabular}
\begin{tabular}{ccccc}
				&3+2							&3+2							&3+1+1								&3+1+1								\\
				&LOW							&HIG							&LOW								&HIG								\\
\hline
$\chi^{2}_{\text{min}}$		&284.4		&256.4		&289.8			&259.0			\\
NDF				&252		&246		&253			&247			\\
GoF				& 8\%		&31\%		& 6\%		&29\%		\\
$\Delta{m}^2_{41}[\text{eV}^2]$	&1.9		&0.93		&1.6			&1.6			\\
$|U_{e4}|^2$			&0.03		&0.015		&0.026			&0.023			\\
$|U_{\mu4}|^2$			&0.012		&0.0097		&0.011			&0.012			\\
$\Delta{m}^2_{51}[\text{eV}^2]$ &4.1		&1.6		&								&								\\
$|U_{e5}|^2$			&0.013		&0.018		&0.0088			&0.0092			\\
$|U_{\mu5}|^2$			&0.0065		&0.0091		&0.0049			&0.0052			\\
$\eta/\pi$			&0.51		&1.6		&0.4			&0.45			\\
\hline
$(\chi^{2}_{\text{min}})_{\text{APP}}$	&87.7	&69.8		&94.8			&75.5		\\
$(\chi^{2}_{\text{min}})_{\text{DIS}}$	&179.1	&179.1		&180.1			&180.1		\\
$\Delta\chi^{2}_{\text{PG}}$	&17.7		&7.5		&14.9			&3.4			\\
$\text{NDF}_{\text{PG}}$	&4		&4		&3			&3			\\
$\text{GoF}_{\text{PG}}$	&0.1\%	&11\%	&0.2\%		&34\%		\\
\hline
$\text{p-val}_{\text{3+1}}$	&$0.12$		&$0.25$		&$0.59$		&$0.42$		\\
$n\sigma_{\text{3+1}}$		&$1.6\sigma$&$1.2\sigma$&$0.54\sigma$	&$0.8\sigma$	\\
\end{tabular}
\end{ruledtabular}
\caption{ \label{tab:3p2}
Results of the fit of short-baseline data
taking into account all MiniBooNE data (LOW) and
only the MiniBooNE data above 475 MeV (HIG)
in the framework of 3+2 and 3+1+1 neutrino mixing.
The notation is similar to that in Tab.~\protect\ref{tab:3p1}.
The last two lines give the p-value ($\text{p-val}_{\text{3+1}}$)
and the corresponding number of excluding $\sigma$'s ($n\sigma_{\text{3+1}}$) of the 3+1 scheme.
}
\end{table}

Let us consider now the fit of short-baseline data
in the framework of 3+2 mixing,
which was considered to be interesting in 2010
when the MiniBooNE neutrino
\cite{0812.2243}
and antineutrino
\cite{1007.1150}
data showed a CP-violating tension.
Unfortunately,
this tension reduced considerably in the final MiniBooNE data
\cite{1303.2588}
and from Tab.~\ref{tab:3p2}
one can see that there is little improvement of the 3+2 fit
with respect to the 3+1 case,
in spite of the four additional parameters and the additional possibility of CP violation.
First,
from Fig.~\ref{fig:MB}
one can see that the 3+2-LOW fit is as bad as the 3+1-LOW fit
in fitting the three anomalous MiniBooNE low-energy bins\footnote{
One could fit the three anomalous MiniBooNE low-energy bins
in a 3+2 scheme \cite{1207.4765}
by considering the appearance data without the
ICARUS \cite{1307.4699}
and
OPERA \cite{1303.3953}
constraints,
but the corresponding relatively large transition probabilities are excluded
by the disappearance data.
}.
Moreover,
comparing Tabs.~\ref{tab:3p1} and \ref{tab:3p2}
one can see that the appearance-disappearance tension
in the 3+2-LOW fit is even worse than that in the 3+1-LOW fit,
since the $\Delta\chi^{2}_{\text{PG}}$ is so much larger that it cannot be compensated
by the additional degrees of freedom
(this behavior has been explained in Ref.~\cite{1302.6720}).
Hence,
as in the 3+1 case it is wise to neglect the three low-energy MiniBooNE anomalous bins
and consider as more reliable the 3+2-HIG fit,
which has an acceptable appearance-disappearance parameter goodness-of-fit.
However,
one must ask if considering the larger complexity of the 3+2 scheme
is justified by the data.
The answer is negative\footnote{
See however the somewhat different conclusions reached in Ref.~\cite{1303.3011}.
}
because,
as one can see from Tab.~\ref{tab:3p2},
the value of the p-value obtained by restricting the 3+2 scheme to 3+1
disfavors the 3+1 scheme only at
$1.2\sigma$
in the 3+2-HIG fit.

A puzzling feature of the 3+2 scheme
is that it needs the existence of two sterile neutrinos
with masses at the eV scale.
We think that it may be considered as more plausible that
sterile neutrinos have a hierarchy of masses.
Hence, we considered also the 3+1+1 scheme
\cite{1010.3970,1201.6662,1205.1791,1306.6079}
in which $m_{5}$ is much heavier than 1 eV
and the oscillations due to
$\Delta{m}^2_{51}$
are averaged.
Hence,
in the analysis of short-baseline data
the 3+1+1 scheme
has one effective parameter less than the 3+2 scheme.
The results of the
3+1+1-LOW and 3+1+1-HIG
fits presented in Tab.~\ref{tab:3p2}
show that the 3+1+1-LOW is as bad as
the 3+1-LOW and 3+2-LOW fits
(see also the bad fit of the three low-energy MiniBooNE anomalous bins
in Fig.~\ref{fig:MB}).
On the other hand,
the 3+1+1-HIG appearance-disappearance parameter goodness-of-fit
is remarkably good,
with a
$\Delta\chi^{2}_{\text{PG}}$
that is smaller than those in the 3+1-HIG and 3+2-HIG fits.
However,
the $\chi^2_{\text{min}}$ in the 3+1+1-HIG is only slightly smaller than that in the
3+1-HIG fit
and the high p-value of the 3+1 scheme
does not allow us to prefer the more complex 3+1+1.

In conclusion,
we have presented the results of the global analysis of all the available
data of short-baseline neutrino oscillation experiments
in the framework of 3+1, 3+2 and 3+1+1 neutrino mixing schemes.
We have shown that the data do not allow us to
reject the simplest 3+1 scheme in favor of
the more complex 3+2 and 3+1+1 schemes.
We have also shown that
the low-energy MiniBooNE anomaly cannot be explained by neutrino oscillations
in any of these schemes.
Considering the preferred 3+1 scheme,
we have updated the constraints on the oscillation parameters
and we have shown that there is only one allowed region in the parameter space
around $\Delta{m}^2_{41} \simeq 1-2 \text{eV}^2$.
We have also shown that the crucial indication
in favor of short-baseline
$\bar\nu_{\mu}\to\bar\nu_{e}$
appearance is still given by the old LSND data
and the MiniBooNE experiment has been inconclusive.
Hence new better experiments are needed in order to
check this signal
\cite{1204.5379,1304.2047,1307.7097,1308.0494}.
Let us finally emphasize that,
besides the direct observation of
short-baseline
$\nua{\mu}\to\nua{e}$
transitions,
it is crucial to observe also
short-baseline
$\nua{e}$ and $\nua{\mu}$
disappearance.
Since the reactor and Gallium anomalies
indicate that $\nua{e}$ indeed disappear,
it is important to search also for the disappearance
of $\nua{\mu}$
\cite{1305.1419,1306.3455}.

\section*{Acknowledgment}

The work of H.W.L. is
supported in part by the National Natural Science Foundation of
China under Grant No.~11265006. The work of Y.F.L. is supported in
part by the National Natural Science Foundation of China under Grants
No.~11135009 and No.~11305193.


\end{document}